\documentclass[doublecol]{epl2} 

\title{{Covariant form of the ideal magnetohydrodynamic  ``connection theorem'' in a relativistic plasma}}
\shorttitle{Covariant connection theorem} 

\author{F. Pegoraro}

\institute{                    
  \inst{} Dipartimento di Fisica ``Enrico Fermi'', Universit\`a di Pisa, Largo B. Pontecorvo 3, I-56127 Pisa, Italy\\
}

\abstract{The magnetic connection theorem of ideal Magnetohydrodynamics  by Newcomb  [Newcomb W.A., {\it Ann. Phys.},  {\bf 3} , 347 (1958)] and its covariant formulation are rederived   and reinterpreted 
in terms of a ``time resetting''  projection that accounts for  the loss of simultaneity in different reference frames between spatially separated events.}

\begin{document}

\maketitle

\section{Introduction}

The dynamics of relativistic plasmas is presently under extensive investigation, both in the context of laboratory plasmas  such as laser produced plasmas  and in the context of high energy astrophysics,  and the concept of relativistic magnetic reconnection is frequently  used in the literature (see e.g., Ref.\cite{ZHK}).   

However, the meaning of   ``magnetic connection''  (i.e.,  of the property that should be broken by the ``reconnection processes''  leading to the release of magnetic energy),  and the concept   of magnetic field line motion  and,  in fact,  of magnetic topology itself may not  be  evident  in the case of a relativistic plasma dynamics where the distinction between electric and magnetic fields is frame dependent.  This is in particular the case in  the presence of large, inhomogeneous velocity fields  where it is not clear how to identify a preferred reference frame.

\section{Magnetic connections of fluid plasma elements}
Within the Magnetohydrodynamic  (MHD) description of a nonrelativistic plasma,  the ideal Ohm's law ${\vec E} + {\vec v} \times {\vec B}  =  0$ allows us to give a meaning to the concept of motion of a  magnetic field line. 

 In fact it can be shown\cite{Newcomb}  that  if  two plasma elements initially  located at  positions $\vec{x}_{1}$ and $\vec{x}_{2}$ are connected by a magnetic line,   and if they move with  the (fluid) plasma velocity ${\vec v}({\vec x},t)$  that satisfies the ideal Ohm's law, then for every following time  there will be a  magnetic line that connects  them.   Here ${\vec B}({\vec x},t) $ and ${\vec E}({\vec x},t)$ are the magnetic and electric fields  in the plasma.  
 
 The proof of this statement  does  not require the use of the ``inhomogeneous''  Maxwell's equations, i.e. of the Maxwell's equations that relate the electromagnetic fields to their sources. Thus it does not depend on the assumptions  that are made in nonrelativistic MHD, such as quasineutrality and the neglecting of the displacement current that cannot be applied to a relativistic MHD plasma description\cite{RMHD}.
In other words, provided the ideal Ohm's law is satisfied, the equation  for the magnetic field  written   in three-dimensional (3D)  notation 
\begin{equation}
{\partial {\vec B}}/ {\partial t}   - \nabla  \times  ( {\vec v } \times  {\vec B} ) = 0 
\label{1} \end{equation} 
is valid   independently of the plasma being relativistic or not, although clearly  it is  not explicitly covariant.

With appropriate  assumptions on the smoothness of the velocity field  the ``connection theorem'' can be  proved by computing the Lagrangian derivative along the fluid trajectories of the expression $\delta \vec{l}\times\vec{B}$ where $\delta \vec{l}$ is the 3D  vector connecting to infinitesimally close fluid elements.
 {We obtain \begin{equation}
\frac{d}{dt}\vec{B}=\vec{\nabla}\times\left(\vec{v}\times\vec{B}\right)+
\left(\vec{v}\cdot\vec{\nabla}\right)\vec{B},  \label{1a} \end{equation} 
 \begin{equation}\frac{d}{dt} \delta \vec{l} =\left(\delta 
\vec{l}\cdot\vec{\nabla}\right)\vec{v}, \label{1b} \end{equation} 
and finally,}
\begin{eqnarray} \frac{d}{dt} {\left(\delta
\vec{l}\times\vec{B}\right)} =  -{\left(\delta
\vec{l}\times\vec{B}\right)}\left(\vec{\nabla}\cdot\vec{v}\right) \nonumber \\ 
-\left[{\left(\delta\vec{l}\times\vec{B}\right)}\times\vec{\nabla}\right]\times\vec{v}, ~~~~~~~~~~~~~  \label{2} \end{eqnarray}  
which ensures that if a $t=0$ the  $\delta \vec{l}$ separation vector is  parallel  to $\vec{B}$, i.e. if  $[\delta \vec{l} \times\vec{B}]_{t=0}= 0$,\,   then 
it remains zero at all times.   {As for Eq.(\ref{1}),  Eqs.(\ref{1a}-\ref{1b}) are valid   independently of the plasma being relativistic or not, although   they are  not explicitly covariant. }

A relativistically covariant generalization of the concept of magnetic topology (see e.g. the extended analysis presented in Ref.\cite{Hornig})   encounters two major related obstacles:
the loss of the concept of simultaneity   between spatially separated events in different reference frames  and the fact that the electric and the magnetic fields are not independent quantities but are the components  of an antisymmetric tensor that transforms in different frames under appropriate Lorentz transformations.

 It is possible however, in line with the results presented in Ref.\cite{Newcomb} but with a different perspective, to rewrite  Eqs.(\ref{1}-§ \ref{2})  { in a fully covariant form 
and to generalize in a frame independent way } the concept of magnetic connection to a relativistic plasma within the ideal MHD description. 

\section{Covariant Lagrangian field equation}

Using standard relativistic notation with Greek indices running from $0$ to $3$ and the summation convention over dummy indices,  the ideal  Ohm's law   in four dimensional (4D) covariant notation  takes the form
 \begin{equation} F_{\mu\nu}u_\mu =0, \label{3} \end{equation}  where $F_{\mu\nu}(x_\alpha) $ is the electromagnetic field tensor\cite{Landau}, $x_\alpha $ is the position  four vector and $u_\alpha = d x_\alpha/ d \tau $ the four velocity  of the fluid element   with $\tau$ its proper time.\\  Equation (\ref{3}) can be rewritten as  a covariant equation  for the variation of the four vector potential $ A_\nu$ along the fluid motion in the (Lagrangian) form
\begin{equation} d A_\nu / d  \tau = u_\mu  (\partial _\nu A_\mu), \label{4} \end{equation} 
where $ d  / d \tau= u_\mu  \partial _\mu $ has been used with $\partial_\mu \equiv \partial /  \partial x_\mu $.
 {By computing the Lagrangian derivative of  the electromagnetic field tensor  with respect to the proper time $\tau$, using  Eq.(\ref{4}) and recalling that $\partial_\mu\,  ( d /d \tau ...)  \not \,  =  \, d /d \tau  \, ( \partial_\mu  ...)$,   we obtain 
\begin{equation}  dF_{\mu\nu}/d  \tau   = (\partial_\mu u_\alpha)  \,  F_{\nu\alpha}  -  (\partial_\nu u_\alpha)  \,  F_{\mu\alpha}, \label{5} \end{equation}   
which is the  covariant counterpart of Eq.(\ref{1}) or, more exactly, of its Lagrangian form given by Eq.(\ref{1a}).}

In a   given frame, projecting Eq.(\ref{5}) onto its space-space and space-time components and  writing  $(\partial_\mu u_\alpha)  \,  F_{\nu\alpha} =  \gamma F_{\nu\alpha} \partial_\mu (u_\alpha /\gamma)$ with $\gamma$ the Lorentz factor (where we use  $F_{\alpha\nu} u_\alpha=0$) we recover Eq.(\ref{1a})  for the magnetic field,   in terms of the Lagrangian time derivative $d/ dt = \partial /\partial t + {\vec v} \cdot  \nabla$,\,  plus  the associated equation for the Lagrangian  time derivative of the electric  field ${\vec E}$. 

\section{Covariant connection equation} 

Let us now introduce  the spacelike event-separation four vector $d l_\mu$  and refer first  to the  frame  where $d l_0 = 0 $,  {i.e., where the two events are simultaneous.}\,\, 
In this frame the condition $F_{\mu\nu}d l_\mu =0 $   is  equivalent to $d {\vec  l} \times {\vec B} =0$ and includes the additional condition  $d {\vec  l} \cdot {\vec E} =0$ that follows from the ideal Ohm's law if $d {\vec  l} \times {\vec B} =0$.

 Note that the condition $F_{\mu\nu}d l_\mu =0 $  requires that  the Lorentz invariant $F_{\mu\nu} F_{\mu\nu}^*$  vanishes  (i.e. ${\vec E}\cdot {\vec B} =0$). Here $F_{\mu\nu}^*  
 \equiv \varepsilon_{\mu\nu\alpha\beta} F_{\alpha \beta} /  2$  is the dual tensor of $F_{\mu\nu}$ and  $\varepsilon_{\mu\nu\alpha\beta}$ is the completely  antisymmetric Ricci tensor in 4D. \,  Because of the ideal Ohm's law (\ref{3}),  this condition is satisfied. 
 
 Then, independently of the frame chosen,  $d l_\mu$ belongs to a 2D hyperplane. If the Lorentz invariant  $F_{\mu\nu} F_{\mu\nu} $ is negative, i.e., if  $E^2< B^2$ (which  is the case we  consider since $u_\mu$ is timelike), in this hyperplane we can choose one timelike (e.g., along $u_\mu$) and one spacelike direction  (along $d l_\mu$).

Defining  the 4D displacement  of a plasma  fluid element $ \Delta x_\mu = u_\mu \Delta\tau$, we find  \begin{equation} \Delta d l_\mu = [d l_\alpha (\partial_\alpha u_\mu)] \Delta \tau +  u_\mu  
  [d l_\alpha (\partial_\alpha \Delta \tau )] ,\label{6} \end{equation}
  { that generalizes the corresponding 3D expression given by Eq.(\ref{1b})  and  includes the coordinate dependence of the proper time variation. }
  
  Finally,  from  Eqs.(\ref{5},\ref{6}), again  using  $F_{\mu\nu} u_\mu=0$, we obtain
\begin{equation} \frac{d  }{ d \tau}  (d l_\mu F_{\mu\nu}  )\, =  \, -(\partial_\nu u_\beta) (d l_\alpha F_{\alpha \beta}).\label{7}\end{equation}
which is the covariant form of Eq.(\ref{2}).

\section{Projection along the trajectories}
 
In a reference frame where $d l_0\not= 0$, i.e., where the two events are not simultaneous,  we can  ``project''  $ d l_\mu$ onto 3D space along the fluid trajectories  by defining   { $d l_\mu^\prime  =   d l_\mu - u_\mu\,   d\lambda,$  such that $d l_0^\prime =0$,} without changing the connection equation since $F_{\mu\nu}u_\mu =0$.
This shows that in order  to recover the standard form of the connection theorem in terms of the magnetic field alone it is sufficient  to {reset the time} by  moving  the endpoints of the  word-line connecting the two close events  along their trajectories. This result  amounts to a rewording of the corresponding discussion given in Ref.\cite{Newcomb}.

\section{Conclusions}

The magnetic connection theorem of ideal MHD can be cast in a covariant, frame independent, form but its interpretation in terms of magnetic field lines alone requires that  the time of the two connected plasma elements be reset so as to restore simultaneity   when the reference frame is changed.   

This procedure is made possible by the assumption that the relativistic plasma dynamics obeys the ideal Ohm's law  that allows us to  move  the endpoints of the  word-line connecting the two close events  along their trajectories without changing the connection equation.

 {Having established a covariant formulation of the connection theorem it will now be possible to reconsider in a covariant framework the related topological properties of an ideal  MHD plasma,  such as in particular the so called ``linking number'' between closed magnetic field lines.  

Independently of this topological formulation we note that Eq.(\ref{5}) represents a convenient form for advecting the e.m.  fields in an ideal MHD plasma in a frame independent way, while Eq.(\ref{7}) can provide an accuracy test for Relativistic ideal-MHD numerical codes. 

More importantly,  the time resetting  procedure obtained  by  a projection along the trajectories of the plasma elements  should allow us to describe in a frame independent way  the redistribution of the magnetic plasma  configuration in the presence of a reconnection event, when the connections are   locally broken in space and time as Eq.(\ref{1}) is  locally violated.  Vice-versa it will allow us  to locate in a frame independent way  where in space and in time the reconnection event  has occurred (see e.g., the point raised in Ref.\cite{zen11}). 
As mentioned before, this  can be important in the presence of large inhomogeneous velocity fields that prevent us from selecting a preferred reference frame.}

\acknowledgments
The research leading to these results has received funding from the European Commission's Seventh
Framework Programme (FP7/2007Ð2013) under the  SWIFF  grant agreement (project n¡ 263340, www.swiff.eu).

\end{document}